\documentclass[]{piparticle-final}
\usepackage{graphicx}
\usepackage{amsmath}
\usepackage{cite} 
\usepackage{epstopdf}

\begin{document}

\volume{4}               
\articlenumber{040005}   
\journalyear{2012}       
\editor{M. C. Barbosa}   
\received{3 September 2012}     
\accepted{24 October 2012}   
\runningauthor{R D Porasso \itshape{et al.}}  
\doi{040005}         

\title{A criterion to identify the equilibration time in lipid bilayer simulations}

\author{Rodolfo D. Porasso,\cite{inst1}\thanks{E-mail: rporasso@unsl.edu.ar}\hspace{0.5em}
        J.J. L\'opez Cascales,\cite{inst2}}

\pipabstract{
With the aim of establishing a criterion for identifying when a lipid bilayer has reached steady state using the molecular dynamics simulation technique, lipid bilayers of different composition in their liquid crystalline phase were simulated in aqueous solution in presence of CaCl$_2$ as electrolyte, at different concentration levels. In this regard, we used two different lipid bilayer systems: one composed by 288 DPPC (DiPalmitoylPhosphatidylCholine) and another constituted by 288 DPPS (DiPalmitoylPhosphatidylSerine). In this sense, for both type of lipid bilayers, we have studied the temporal evolution of some lipids properties, such as the surface area per lipid, the deuterium order parameter, the lipid hydration and the  lipid-calcium coordination. From their analysis, it became evident how each property has a different time to achieve  equilibrium.  The following order was found, from faster property to slower property: coordination of ions $\approx$ deuterium order parameter $>$ area per lipid $\approx$ hydration. Consequently, when the hydration of lipids or the mean area per lipid are stable, we can ensure that the lipid membrane has reached the steady state.
}

\maketitle

\blfootnote{
\begin{theaffiliation}{99}
   \institution{inst1} Instituto de Matem\'{a}tica Aplicada San Luis (IMASL) - Departamento de F\'{i}sica, Universidad Nacional de San Luis/CONICET, D5700HHW, San Luis, Argentina.
   \institution{inst2} Universidad Polit\'{e}cnica de Cartagena, Grupo de Bioinform\'{a}tica y Macromol\'{e}culas (BioMac) Aulario II, Campus de Alfonso XIII, 30203 Cartagena, Murcia, Spain.
\end{theaffiliation}
}

\section{Introduction}
Over the last few decades, different computational techniques have emerged in different fields of science,  some of them being extensively implemented and used by a great number of scientists around the globe. Among others, the Molecular Dynamics (MD) simulation is a very popular computational technique, which is widely used to obtain insight with atomic detail of steady and dynamic properties in the fields of biology, physics and  chemistry. In this regard, a critical aspect that must be identified in all the MD simulations is related  to the required equilibration time to achieve a steady state. This point is crucial in order to avoid simulation artifacts that could lead to wrong conclusions. Currently, with the increment of the computing power accessible to different investigation groups, much longer simulation trajectories are being carried out to obtain reliable information about the systems, with the purpose of approaching the time scale 
of the experimental phenomena. However, even when 
this fact is objectively desirable without further objections, nowadays, much longer equilibration times are arbitrarily being required by certain reviewers during the revision process.  From our viewpoint, this should be thoroughly revised due to the following two main reasons: first, because it results in a limiting factor  in the use of this technique by other  research groups which cannot access to very expensive computing centers (assuming that authors provide enough  evidence of the equilibration of the system).  Second, to avoid wasting expensive computing time in the study of certain properties which do not require such long equilibration times, once the steady state of the system has been properly identified.

Phospholipid bilayers are of a high biological relevance, due to the fact that they play a crucial role in the control of the diffusion of small molecules, cell recognition, and signal transduction, among others. In our case, we have chosen the PhosphatidylCholine (PC) bilayer because it has been very well studied by MD simulations \cite{berger:97,sjheb94,essmann:95b,feller:95,shinoda:95,tieleman:96,tu:95} and experimentally as well \cite{brown:84,brown:82,nagle:96,rand:89,seelig:77,seelig:80,sun:94}. Furthermore, studies of the effects of different types of electrolytes on a PC bilayer have also been studied, experimentally \cite{akutsu:81,ganesan:82,herbette:84,huster:00,inoko:75,lehrmann:94,lis:81b,lis:81a,shibata:90,tatulian:91} and by  simulation \cite{bockmann:04,faraudo:07,gurtovenko:05,tieleman:04,pandit:03,pedersen:06,woolf:04,mikami:07,yamada:05}.

As  mentioned above, the Molecular Dynamics (MD) simulations have emerged during the last decades as a powerful tool to obtain insight with atomic detail of the structure and dynamics of lipid bilayers \cite{smit:02,jjlc:96a,berendsen:90}. Several MD simulations of membranes under the influence of different salt concentrations have been carried out. One of the main obstacles related to these studies has been the time scale associated to the binding process of ions to the lipid bilayer. Considering the literature, a vast dispersion of equilibration times associated to the binding of ions to the membrane  has been reported, where values ranging from 5 to 100 ns have been suggested for monovalent and divalent cations \cite{bockmann:04,gurtovenko:05,tieleman:04,pandit:03,mikami:07,bockmann:03,gurtovenko:08}. In this regard, we carried out four independent simulations of a lipid bilayer formed by 288 DPPC in aqueous solutions, for different concentrations of CaCl$_2$  to 
provide an overview of their equilibration 
times. Among other properties, the surface area per lipid, the deuterium order parameters, lipid hydration and lipid-calcium coordination were studied. 

Finally, in order to generalize our results, a bilayer formed by 288 DPPS  in its liquid crystalline phase, in presence of CaCl$_2$ at 0.25N, was simulated as well .

\section{Methodology}

Different molecular Dynamics (MD)  simulations of lipid bilayer formed by 288 DPPC were carried out in aqueous solutions for different concentrations of CaCl$_2$, from  0, up to 0.50 N. Furthermore, with the aim of generalizing our results, a bilayer of 288 DPPS in presence of CaCl$_2$ at 0.25 N was simulated as well. Note that the concentration of CaCl$_2$ in terms of normality is defined as:

\begin{equation}
\textrm{normality}= \frac{n_\textrm{equivalent grams}}{l_\textrm{solution}}
\label{eq_normal}
\end{equation}
where $n_\textrm{equivalent grams}=\frac{\textrm{gr(solute)}}{\textrm{equivalent weight}}$ and $\textrm{equivalent weight} = \frac{\textrm{Molecular weight}}{n}$, being $n$ the charge of the ions in the solution.

In Table \ref{table_simulation},  the number of molecules that constitute each system, applying Eq. (\ref{eq_normal}), is summarized. 

\begin{table}
  {\begin{tabular}{llllll} 
\small Type of Lipid & \small [CaCl$_2$] N & \small Ca$^{2+}$ & \small Cl$^-$ & \small Water \\ \hline \hline
 DPPC & 0     & 0 & 0 & 10068 \\ 
 DPPC & 0.06 &  5 & 10 & 10053\\ 
 DPPC & 0.13 & 12 & 24 & 10032\\ 
 DPPC & 0.25 & 23 & 46 & 9999\\ 
 DPPC & 0.50 &46 & 92 & 9930\\
 DPPS & 0.25 & 204  & 120 & 26932\\  \hline \hline
\end{tabular}}
\caption{The simulated bilayer systems. Note that the salt concentration is given in normal units. The  numerals describe the number of molecules contained in the simulation box.}
\label{table_simulation}
\end{table}

\begin{figure}
\begin{center}
\includegraphics[width=0.45\textwidth]{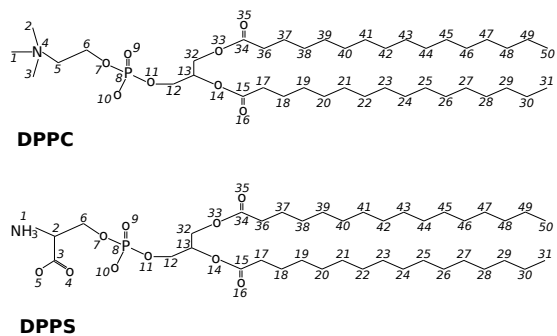}
\end{center}
\caption{Structure and atom numbers for DPPC and DPPS used in this work.}\label{fig_dppc}
\end{figure}

To build up the original system, a single DPPC lipid molecule, or DPPS lipid (Fig. \ref{fig_dppc}), was placed  with its molecular axis perpendicular to the membrane surface ($xy$ plane). Next, each DPPC, or DPPS, was randomly rotated and  copied 144 times on each leaflet of the bilayer.  Finally, the gaps existing in the computational box (above and below the phospholipid bilayer) were filled using an equilibrated box containing 216 water molecules of the extended simple point charge (SPC/E) \cite{spce} water model.

Thus, the starting point of the first system of Table \ref{table_simulation} was formed by 288 DPPC in absence of CaCl$_2$. Once  this first system was generated, the whole system was subjected to the steepest descent minimization process to remove any excess of strain associated  with overlaps between neighboring atoms of the system. Thereby, the following DPPC systems in presence of CaCl$_2$ were generated as follows: to obtain a [CaCl$_2]$=0.06 N, 15 water molecules were randomly substituted by 5 Ca$^{2+}$ and 10 Cl$^-$. An analogous procedure was applied to the rest of the systems, where 36, 69 and 138 water molecules were substituted by 12, 23 and 46 Ca$^{2+}$ and 24, 46 and 92 Cl$^-$, to obtain a [CaCl$_2]$ concentration of 0.13 N, 0.25 N and 0.50 N, respectively. Finally, the DPPS bilayer was generated following the same procedure described above for the DPPC, starting from a single DPPS molecule and once the lipid bilayer in presence of water passed the minimization 
process, 324 water molecules were  
substituted by 204 Ca$^{2+}$ and 120 Cl$^-$ (note that 144 of the 204 calcium ions were added to balance the negative charge associated  with the DPPS).

GROMACS 3.3.3 package \cite{gromacs:95,gromacs:01} was used in the simulations, and the properties showed in this work were obtained using our own code. The force field proposed by Egberts et al. \cite{sjheb94} was used for the lipids, and a time step of 2 fs was used as integration time in all the simulations. A cut-off of 1.0 nm was used for calculating the Lennard-Jones interactions. The electrostatic interaction was evaluated using the particle mesh Ewald method \cite{darden:93,essmann:95}. The real space interaction was evaluated using a 0.9 nm cut-off, and the reciprocal space interaction using a 0.12 nm grid with a fourth-order spline interpolation. A semi-isotropic coupling pressure was used for the coupling pressure bath, with a reference pressure of 1 atm which allowed the fluctuation of each axis of the computer box independently. For the DPPC bilayer, each component of the system (i.e., lipids, ions and water) was coupled to an external temperature coupling bath at 330 K, which is well 
above the 
transition temperature of 314 K \cite{young:88,seelig:74}.  For DPPS bilayer, each component of the system was coupled to an external temperature coupling bath at 350 K, which is above the transition temperature \cite{cevcbio81,hebio82}.  All the MD  simulations were carried out using periodic boundary conditions. The total trajectory length of each simulated system was of 80 ns of MD simulation, where the coordinates of the system were recorded every 5 ps for their appropriate analysis.

Finally, in order to study the effect of the temperature, only the case corresponding to 0.25 N CaCl$_2$ was investigated at two additional temperatures, 340 K and 350 K.

\section{Results and discussion}
\subsection{Effect of the CaCl$_2$ concentration}
\subsubsection{Surface area per lipid}
Surface area per lipid $\langle A \rangle$ is a property of lipid bilayers which has been accurately measured from experiments \cite{nagle:00}. The calculation of mean area per lipid can be determined from the MD simulation as:

\begin{equation}
 \langle A\rangle = \frac{x \cdot y}{N}
\label{eq_area}
\end{equation}
where $x$ and $y$ represent the box sizes in the direction $x$ and $y$ (perpendicular to the membrane surface) over the simulation, and $N$ is the number of lipids contained in one leaflet, in our case $N=144$.

\begin{figure}[b]
\begin{center}
\includegraphics[width=0.45\textwidth]{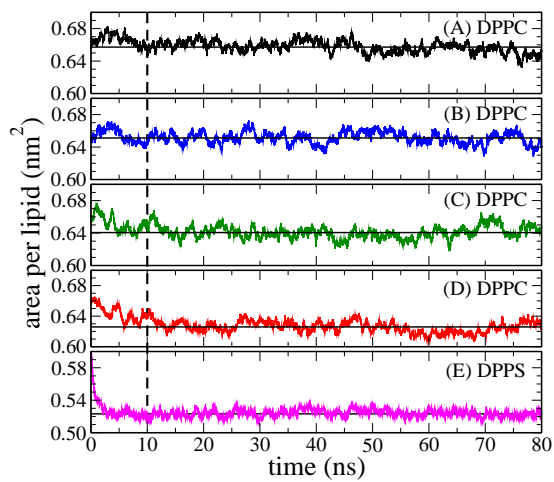}
\end{center}
\caption{Running area per lipid at T = 330 K in presence of [CaCl$_2$] at (A) 0.06 N, (B) 0.13 N, (C) 0.25 N, (D) 0.50 N and (E) 0.25 N (in this case, T = 350 K). Solid lines represent the mean area obtained from the last 70 ns of the simulated trajectories (see text for further explanation).  The type of lipid is indicated in the legends.} \label{fig_area_T_330}
\end{figure}

Focusing on the study of the time evolution of the area per lipid, Figure \ref{fig_area_T_330} depicts the running surface area per lipid for different concentrations of CaCl$_2$ and type of lipid. In general, for the 5 bilayers formed by DPPC or DPPS, the area per lipid achieved a steady state after 10 ns of simulation, being this equilibration time almost independent of the concentration of CaCl$_2$ and type of lipid which composed the membrane.

In absence of salt, an average area per lipid of $\langle A \rangle = 0.663 \pm 0.008$ nm$^2$ was calculated from the last 70 ns of the simulated trajectory, discarding the first 10 ns corresponding to the equilibration time. This value agrees with experimental data, where values in a range from 0.55 to 0.72 nm$^2$ have been measured \cite{nagle:96,rand:89,nagle:00,lewis:83,pace:82,thurmond:91}. Table \ref{table_summary} shows the mean surface area per lipid (again, after discarding the equilibration time of 10 ns) with their corresponding error bar.  

From the simulation results,  a shrinking in the  surface area per lipid with the increment of the ionic strength of the solution is observed. This shrinking is expected and attributed to the complexation of lipid molecules by calcium, such as it has been pointed out in previous studies \cite{tieleman:04,pandit:03,porasso:09}.

\subsubsection{Deuterium order parameter}
The deuterium order parameter, $S_{CD}$, is measured from $^2$H-NMR experiments. This parameter  provides relevant information related  to the disorder of the hydrocarbon region in the interior of the lipid bilayers by measuring the orientation of the hydrogen dipole of the methylene groups with respect to the perpendicular axis to the lipid bilayer. Due to the fact that hydrogens of the lipid methylene groups (CH$_2$) have not been taken into account (in an explicit way) in our simulations, the order parameter $-S_{CD}$ on the $i + 1$ methylene group was defined as the normal unitary vector to the vector defined from the $i$ to the $i +2$ CH$_2$ group and contained in the plane formed by the methylene groups $i$, $i + 1$ and $i + 2$. Thus, the deuterium order parameter $-S_{CD}$ on the $i-th$ of the CH$_2$ group can be estimated by Molecular Dynamics  simulations as follows:

\begin{equation}
-S_{CD}=\frac{1}{2} \langle{3\cos^{2}(\theta)-1}\rangle
\label{eq_SCD}
\end{equation}

where $\theta$ is the angle formed between the unitary vector defined above and the $z$ axis. The expression in  brackets $\langle \dots \rangle$ denotes an average over all the lipids and time. Hence, note that the $-S_{CD}$ can adopt any value between -0.5 (corresponding to a parallel orientation to the lipid/water interface) and 1 (oriented along the normal axis to the lipid bilayer).

\begin{table}
{\begin{tabular}{clll} 
\small Type of & \small [CaCl$_2$] N &  \small $\langle A \rangle$ (nm$^2$) & \small Hydration \\
\small Lipid &  &  & \small Number \\ \hline \hline
DPPC & 0    & \small 0.663 $\pm$0.008 & \small 1.758 $\pm$0.009 \\ \hline
DPPC & 0.06 & \small 0.658 $\pm$0.008 & \small 1.740 $\pm$0.009 \\ 
DPPC & 0.13 & \small 0.651 $\pm$0.007 & \small 1.719 $\pm$0.010 \\ 
DPPC & 0.25 & \small 0.641 $\pm$0.009 & \small 1.680 $\pm$0.015 \\ 
DPPC &0.50 & \small 0.628 $\pm$0.010 & \small 1.610 $\pm$0.015 \\
DPPS & 0.25 & \small 0.522 $\pm$0.007 & \small 2.552 $\pm$0.010 \\ \hline \hline
\end{tabular}}
\caption {Area per lipid and lipid  hydration number as a function of salt concentration (see text for further explanation). Note that the salt concentration is given in normal units. Error bars were calculated for each system separately from subtrajectories of 10 ns length. Simulation temperature = 330 K.} \label{table_summary}
\end{table}

\begin{figure}
\begin{center}
\includegraphics[width=0.45\textwidth]{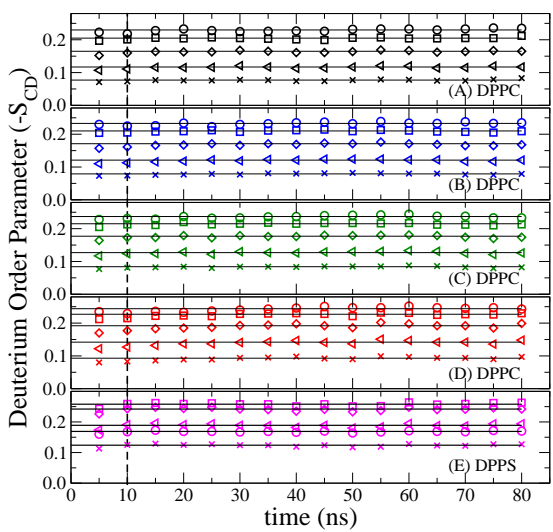}
\end{center}
\caption{Running deuterium order parameter, $-S_{CD}$, in presence of [CaCl$_2$] at (A) 0.06 N, (B) 0.13 N, (C) 0.25 N, (D) 0.50 N and (E) 0.25 N.  DPPC simulations were performed at 330 K and  DPPS simulation temperature was 350 K. Solid lines represent mean values $-S_{CD}$ obtained from the last 70 ns of the simulated trajectories.  The type of lipid is indicated in the legends. Symbols: $\circ$ hydrocarbon 2; $\diamond$ hydrocarbon 6; $\triangleleft$ hydrocarbon 10; $+$ hydrocarbon 13 and $\times$ hydrocarbon 15. Note that the error bars have the same size as symbol.} \label{fig_order_T_330}
\end{figure}

Figure \ref{fig_order_T_330} shows the running $-S_{CD}$ for different carbons of the DPPC and DPPS tails  and salt concentrations. Only the carbons which correspond to the initial (hydrocarbons 2 and 6), the middle (hydrocarbon 10) and final (hydrocarbons 13 and 15) methylene groups of the lipid tails were depicted in this  figure. Each point of the figure  represents the average values of $-S_{CD}$ over 5 ns of subtrajectory length, and the lines represent the mean values calculated from the last 70 ns of the trajectories simulated. From this figure, it is observed how in all the cases, the required equilibration time is less than 10 ns of simulation time, independently of the salt concentration and the type of lipid. Finally, it is noted that Figure \ref{fig_order_T_330}  exhibits an increase in the deuterium order parameters with the salt concentration, consistent with the shrinking of the area per lipid described above.

\subsubsection{Lipid hydration}
To analyze the lipid hydration, the radial distribution function $g(r)$ of water around one of the oxygens of the phosphate group (atom number 10 in Fig. \ref{fig_dppc} for DPPC and DPPS)  was calculated. The radial distribution function $g\left( r \right)$ is defined as follows:

\begin{equation}
g(r)=\frac{N(r)}{4{\pi}r^2{\rho}\delta r}
\label{eq_gr}
\end{equation}
where $N\left( r \right)$ is the number of atoms in a spherical shell at distance $r$ and thickness $\delta r$ from a reference atom.  $\rho$ is the density number taken as the ratio of atoms to the volume of the total computing box.

\begin{figure}
\begin{center}
\includegraphics[width=0.45\textwidth]{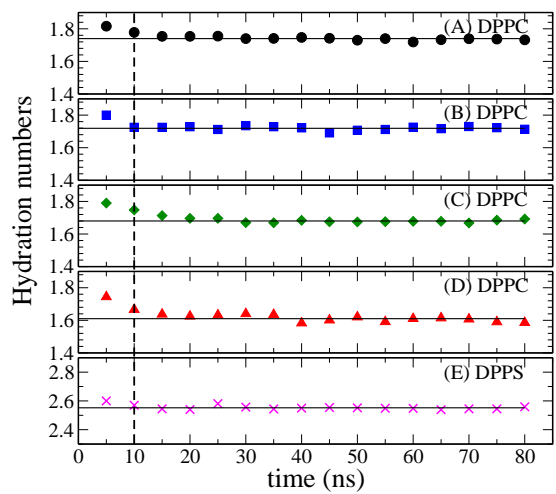}
\end{center}
\caption{Hydration number of the phosphate oxygen (atom 10 in Fig. \ref{fig_dppc}) along the simulated trajectories in presence of [CaCl$_2$] at (A) 0.06 N, (B) 0.13 N, (C) 0.25 N, (D) 0.50 N (for DPPC T = 330 K) and (E) 0.25 N.  In this case, T = 350 K. Solid lines represent the mean value of the hydration number calculated from the last 70 ns of the simulated trajectories.  The type of lipid is indicated in the legends. Note that the error bars have the same size as symbol.} \label{fig_hydration_T_330}
\end{figure}

From numerical integration of the first peak of the radial distribution function, the hydration numbers can be estimated for different atoms of the DPPC or DPPS. Figure \ref{fig_hydration_T_330} depicts the hydration number of phosphate oxygen (atom 10 in Fig. \ref{fig_dppc} for DPPC and DPPS) in presence of CaCl$_2$, where each point represents the average of 5 ns subtrajectory length.  These results show how this property reached a steady state for the cases (A), (B) and (E), after 10 ns of simulation. However, for the cases (C) and (D), 5 ns of extra simulation trajectory were required to reach a steady state. Table \ref{table_summary} shows the hydration numbers for the last 70 ns of the trajectory length, corresponding to four concentrations of CaCl$_2$ and  both  types of lipids, DPPC and DPPS. In this regard, from Fig. \ref{fig_hydration_T_330},  the significant lipid 
dehydration with the increment of the 
ionic strength of the solution is evident, in good accordance with previous 
results \cite{porasso:09}.

\subsubsection{Phospholipid-calcium coordination}
\begin{figure}
\begin{center}
\includegraphics[width=0.45\textwidth]{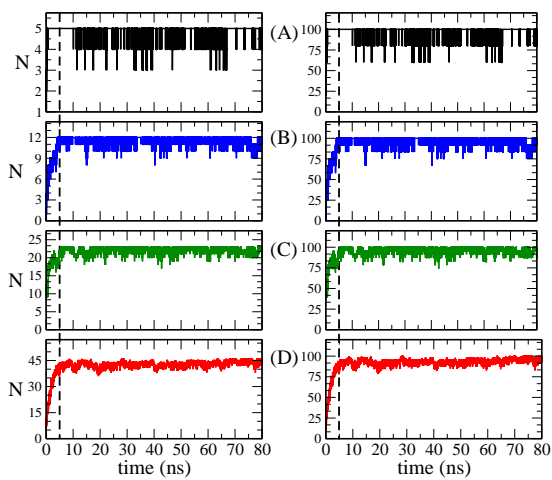}
\end{center}
\caption{Left column represents the number of  Ca$^{2+}$ coordinated to lipids in presence of [CaCl$_2$] at (A) 0.06 N, (B) 0.13 N, (C) 0.25 N and (D) 0.50 N, for T = 330 K. Right column shows the quantity of calcium ions coordinated to lipids expressed in percentage, along the simulated trajectory.} \label{fig_coordination_T_330}
\end{figure}

Some authors have reported how the  lipid coordination by divalent cations widely varies . Thus, on the one hand, some authors \cite{bockmann:04} have reported that this is a very slow process, which requires about 85 ns of simulation time, but, on the other hand, other authors \cite{faraudo:07} have suggested that  this process results much more rapid, taking less than 1 ns. In this sense, the coordination of DPPC-Ca$^{2+}$ was studied by monitoring the oxygen-calcium coordination of the carbonyl oxygens (atoms 16 and 35  in Fig. \ref{fig_dppc}) and phosphate oxygens (atoms 9 and 10 of DPPC  in Fig. \ref{fig_dppc}), as a function of time. The left column  in Fig. \ref{fig_coordination_T_330} represents the oxygen coordination number,  while the right one  depicts the percentage of calcium ions involved in the coordination process with respect to 
the total number of calcium ions present in the 
aqueous solution. Figure \ref{fig_coordination_T_330} shows  how the DPPC coordination by calcium is a quick process, taking less 
than  5 ns of simulation time to achieve a steady state. The kinetic of this process appears to be related  to the ratio between calcium/lipid. Thus, after the first 5 ns of simulation time, the Ca--lipid  coordination presents some fluctuation along the rest of the simulated trajectory. However, in Fig. \ref{fig_coordination_T_330} (A) and (B) (for the cases of lower concentration), it is observed how the percentage of coordination fluctuates between a 60\% and a 100\%. We consider that this broad fluctuation is related to the limited sample size of our simulations  that introduces a certain noise in our results. 

\subsection{Effect of temperature}
This section  focuses on the study of the role played by  temperature on the equilibration process. In this regard, only the system corresponding to a concentration of 0.25 N in CaCl$_2$ was studied, for a range of temperatures from 330 K to 350 K (all of them above the transition temperature of 314 K \cite{young:88,seelig:74} for the DPPC). 

\begin{figure}[h]
\begin{center}
\includegraphics[width=0.45\textwidth]{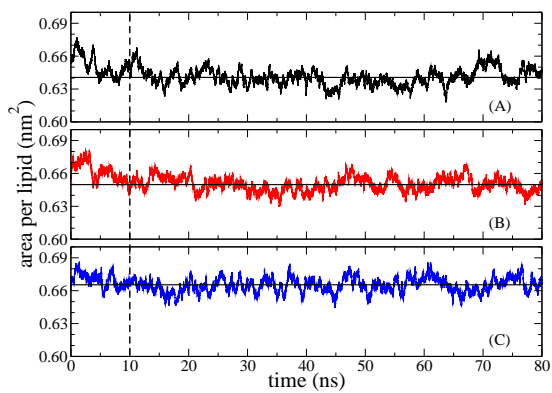}
\end{center}
\caption{Running area per lipid for [CaCl$_2$] = 0.25 N at different temperatures, (A)  T = 330 K, (B) T = 340 K and (C) T = 350 K. Solid lines represent the mean values obtained from the last 70 ns of simulation.} \label{fig_area_Ca_25}
\end{figure}

\begin{figure}[h]
\begin{center}
\includegraphics[width=0.45\textwidth]{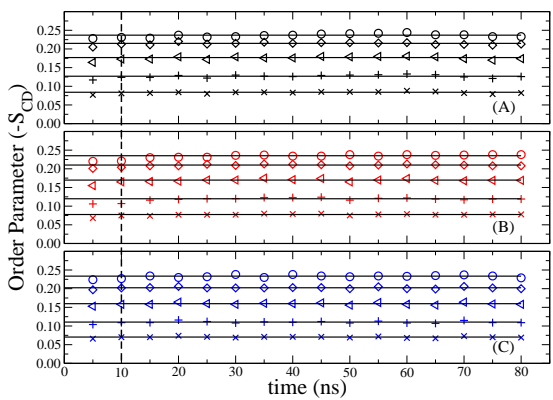}
\end{center}
\caption{Deuterium Order Parameter, $-S_{CD}$, along the simulated trajectory for a concentration of [CaCl$_2$] = 0.25 N,  for the following temperatures: (A) T = 330 K, (B) T = 340 K and (C) T = 350 K. Solid lines represent the average order parameter for the last 70 ns of simulation. Note that the error bars have the same size as symbol.} \label{fig_order_Ca_25}
\end{figure}

\begin{figure}[h]
\begin{center}
\includegraphics[width=0.45\textwidth]{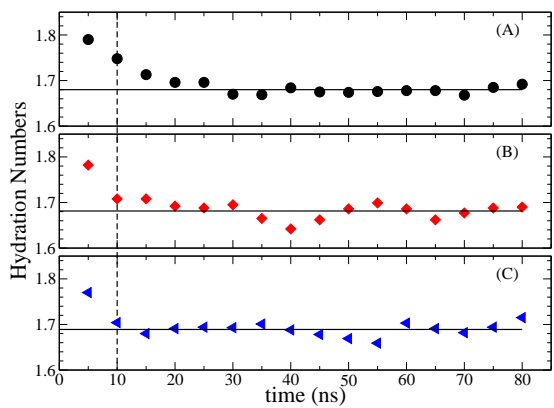}
\end{center}
\caption{Hydration number of phosphate oxygen (atom 10  in  Fig. \ref{fig_dppc}) along the simulated trajectories  for a [CaCl$_2$] = 0.25 N  at different temperatures: (A) T = 330 K, (B) T = 340 K and (C) T = 350 K. Solid lines represent the average hydration number for the last 70 ns of simulation. Note that the error bars have the same size as symbol.} \label{fig_hydration_Ca_25}
\end{figure}

\begin{figure}[h]
\begin{center}
\includegraphics[width=0.45\textwidth]{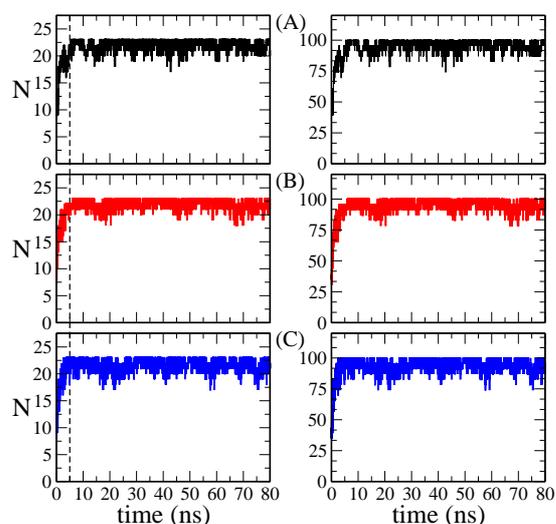}
\end{center}
\caption{Left column represents the number of calcium ions involved in the lipid coordination along time for a concentration of  [CaCl$_2$]= 0.25 N at different temperatures: (A) T = 330 K, (B) T = 340 K and (C) T = 350 K. The right column shows the same information expressed as a percentage of the total number of calcium ions in solution.} \label{fig_coordination_Ca_25}
\end{figure}

Figure \ref{fig_area_Ca_25} shows the running area along the trajectory. In this case, it was noticed how the systems achieve a steady state after a trajectory length of roughly 10 ns, where  Table \ref{table_summary_T} shows the mean area per lipid calculated from the last 70 ns of simulation time. Figure \ref{fig_order_Ca_25} shows the deuterium order parameter of the methylene groups along the lipid tails, calculated from Eq. (\ref{eq_SCD}). Figure \ref{fig_order_Ca_25}, on the one hand, clearly shows that for the three temperatures the systems have reached the steady state before the first 10 ns of simulation.  On the other hand, it shows an increase in the disorder of the lipid tails with temperature, which is closely related with the increase of the area per lipid, such as it was pointed out above. Figure \ref{fig_hydration_Ca_25} depicts the results of the hydration numbers of DPPC for the three temperatures 
studied, where the equilibrated 
state was achieved after 10 ns of simulation time. Table \ref{table_summary_T} provides the calculated hydration numbers in the equilibrium, showing how the lipid hydration remained invariable with the rising of the temperature. Concerning the lipid-calcium coordination, Fig. \ref{fig_coordination_Ca_25} represents the lipid-calcium coordination number, and the right column represents the calcium that  participates in the coordination expressed in percentage respect the total of calcium ions in solution. From simulation, it becomes evident how calcium ions required less than 5 ns to achieve an equilibrated state for the three temperatures studied. In summary, for all the properties studied in this section, a slight decrease in the equilibration time with the increasing temperature was appreciated.

\begin{table}
{\begin{tabular}{lll}
T (K) &  $\langle A \rangle$ (nm$^2$) & Hydration Number \\ \hline
330 & 0.642 $\pm$0.009 & 1.680 $\pm$ 0.010 \\ 
340 & 0.650 $\pm$0.007 & 1.683 $\pm$ 0.020 \\ 
350 & 0.666 $\pm$0.008 & 1.689 $\pm$ 0.015 \\ \hline
\end{tabular}}
\caption{Area per lipid and the lipid hydration number as a function of temperature. Error bars were calculated from subtrajectories of 10 ns length. DPPC bilayer in presence of [CaCl$_2$] = 0.25 N.}\label{table_summary_T}
\end{table}

\section{Conclusions}
The present work deals with the simulation time required to achieve the  steady state for a lipid bilayer system in presence of CaCl$_2$. In this regard, we studied two different systems: one with DPPC and another one with DPPS bilayer; both systems in presence of CaCl$_2$ (at different level concentration).  The salt free case was also studied, as control. The analysis of various lipid properties studied here indicates that some properties reach the  steady state more quickly than  others. In this sense, we found that the area per lipid and the hydration number are slower than the deuterium order parameter and the coordination of cations. Consequently, to ensure that a system composed by a lipid bilayer has reached a  steady state, the criterion that we propose is to show that the area per lipid or the hydration number have reached the equilibrium.

From our results, two important aspects should be remarked:
\begin{enumerate}
 \item
The equilibration time is strongly dependent  on the starting conformation of the system. Wrong starting conformations will require much longer equilibration times, even of one order of magnitude higher than the requested from a more refined starting conformation.
\item
Temperature is a critical parameter for reducing the equilibration time in our simulations, due to the fact that higher temperatures increase the kinetic processes, i.e., the sampling of the configurational space of the system.
\end{enumerate}

\begin{acknowledgements}
Authors wish to thank the assistance of the Computing Center of the Universidad Polit\'ecnica de Cartagena (SAIT), Spain. RDP is member of \textquoteleft Carrera del Investigador\textquoteright, CONICET, Argentine.
\end{acknowledgements}


\begin{thebibliography}{52} 

\bibitem{berger:97} O Berger, O Edholm, F Jahnig, \textit{Molecular dynamics simulations of a fluid bilayer of dipalmitoylphosphatidylcholine at full hydration, constant pressure, and constant temperature}, Biophys. J. \textbf{72}, 2002 (1997).

\bibitem{sjheb94} E Egberts, S J Marrink, H J C Berendsen, \textit{Molecular dynamics simulation of a phospholipid membrane}, Eur. Biophys. J. \textbf{22}, 423 (1994).

\bibitem{essmann:95b} U Essmann, L Perera, M L Berkowitz, \textit{The origin of the hydration interaction of lipid bilayers from MD simulation of dipalmitoylphosphatidylcholine membranes in gel and crystalline phases}, Langmuir \textbf{11}, 4519 (1995).

\bibitem{feller:95} S E Feller, Y Zhang, R W Pastor, R B Brooks, \textit{Constant pressure molecular dynamics simulations: The Langevin piston method}, J. Chem. Phys. \textbf{103}, 4613 (1995).

\bibitem{shinoda:95} W Shinoda, T Fukada, S Okazaki, I Okada, \textit{Molecular dynamics simulation of the dipalmitoylphosphatidylcholine (DPPC) lipid bilayer in the fluid phase using the Nosr-Parrinello-Rahman NPT ensemble}, Chem. Phys. Lett. \textbf{232}, 308 (1995).

\bibitem{tieleman:96} D P Tieleman, H J C Berendsen, \textit{Molecular dynamics simulations of a fully hydrated dipalmitoylphosphatidylcholine bilayer with different macroscopic boundary conditions and parameters}, J. Chem. Phys. \textbf{105}, 4871 (1996).

\bibitem{tu:95} K Tu, D J Tobias, M L Klein, \textit{Constant pressure and temperature molecular dynamics simulation of a fully hydrated liquid crystal phase dipalmitoylphosphatidylcholine bilayer}, Biophys. J. \textbf{69}, 2558 (1995).

\bibitem{brown:84} M F Brown, \textit{Theory of spin-lattice relaxation in lipid bilayers and biological membranes. Dipolar relaxation}, J. Chem. Phys. \textbf{80}, 2808 (1984).

\bibitem{brown:82} M F Brown, \textit{Theory of spin-lattice relaxation in lipid bilayers and biological membranes. $^2H$ and $^{14}N$ quadrupolar relaxation}, J. Phys. Chem. \textbf{77}, 1576 (1982).

\bibitem{nagle:96} J F Nagle, R Zang, S Tristam-Nagle, W S Sun, H I Petrache, R M Suter, \textit{X-ray structure determination of fully hydrated L. phase dipalmitoylphosphatidylcholine bilayers}, Biophys. J. \textbf{70}, 1419 (1996).

\bibitem{rand:89}  R P Rand, V A Parsegian, \textit{Hydration forces between phospholipid bilayers}, Biochim. Biophys. Acta \textbf{988}, 351 (1989).

\bibitem{seelig:77} J Seelig, \textit{Deuterium magnetic resonance: Theory and application to lipid membranes}, Q. Rev. Biophys. \textbf{10}, 353 (1977).

\bibitem{seelig:80} J Seelig, A Seelig, \textit{Lipid conformation in model membranes and biological systems}, Q. Rev. Biophys. \textbf{13}, 19 (1980).

\bibitem{sun:94} W J Sun, R M Suter, M A Knewtson, C R Worthington, S Tristram-Nagle, R Zhang, J F Nagle, \textit{Order and disorder in fully hydrated unoriented bilayers of gel phase dipalmitoylphosphatidylcholine}, Phys. Rev. E. \textbf{49}, 4665 (1994).

\bibitem{akutsu:81} H Akutsu, J Seelig, \textit{Interaction of metal ions with phosphatidylcholine bilayer membranes}, Biochemistry \textbf{20}, 7366 (1981).

\bibitem{ganesan:82} M G Ganesan, D L Schwinke, N Weiner, \textit{Effect of Ca$^{2+}$ on thermotropic properties of saturated phosphatidylcholine liposomes}, Biochim. Biophys. Acta \textbf{686}, 245 (1982).

\bibitem{herbette:84} L Herbette, C A Napolitano, R V McDaniel, \textit{Direct determination of the calcium profile structure for dipalmitoyllecithin multilayers using neutron diffraction}, Biophys. J. \textbf{46}, 677 (1984).

\bibitem{huster:00} D Huster, K Arnold, K Gawrisch, \textit{Strength of Ca$^{2+}$ binding to retinal lipid membrane: Consequences for lipid organization}, Biophys. J. \textbf{78}, 3011 (2000).

\bibitem{inoko:75} Y Inoko, T Yamaguchi, K Furuya, T Mitsui, \textit{Effects of cations on dipalmitoyl phosphatidylcholine/cholesterol/water systems}, Biochim. Biophys. Acta \textbf{413}, 24 (1975).

\bibitem{lehrmann:94} R Lehrmann, J J Seelig, \textit{Adsorption of Ca$^{2+}$ and La$^{3+}$ to bilayer membranes: Measurement of the adsorption enthalpy and binding constant with titration calorimetry}, Biochim. Biophys. Acta \textbf{1189}, 89 (1994).

\bibitem{lis:81b} L J Lis, W T Lis, V A Parsegian, R P Rand, \textit{Adsorption of divalent cations to a variety of phosphatidylcholine bilayers}, Biochemistry \textbf{20}, 1771 (1981).

\bibitem{lis:81a} L J Lis, V A Parsegian, R P Rand, \textit{Binding of divalent cations to dipalmitoylphosphatidylcholine bilayers and its effect on bilayer interaction}, Biochemistry \textbf{20}, 1761 (1981).

\bibitem{shibata:90} T Shibata, \textit{Pulse NMR study of the interaction of calcium ion with dipalmitoylphosphatidylcholine lamellae}, Chem. Phys. Lipids. \textbf{53}, 47 (1990).

\bibitem{tatulian:91} S A Tatulian, V I Gordeliy, A E Sokolova, A G Syrykh, \textit{A neutron diffraction study of the influence of ions on phospholipid membrane interactions}, Biochim. Biophys. Acta \textbf{1070}, 143 (1991).

\bibitem{bockmann:04} R A B\"ockmann, H Grubm\"uller, \textit{Multistep binding of divalent cations to phospholipid bilayers: A molecular dynamics study}, Angewandte Chemie \textbf{43}, 1021 (2004).

\bibitem{faraudo:07} J Faraudo, A Travesset, \textit{Phosphatidic acid domains in membranes: Effect of divalent counterions}, Biophys. J. \textbf{92}, 2806 (2007).

\bibitem{gurtovenko:05} A A Gurtovenko, \textit{Asymmetry of lipid bilayers induced by monovalent salt: Atomistic molecular-dynamics study}, J. Chem. Phys. \textbf{122}, 244902 (2005).

\bibitem{tieleman:04} P Mukhopadhyay, L Monticelli, D P Tieleman, \textit{Molecular dynamics simulation of a palmitoyl-oleoyl phosphatidylserine bilayer with Na$^+$ counterions and NaCl}, Biophys.  J. \textbf{86}, 1601 (2004).

\bibitem{pandit:03} S A Pandit, D Bostick, M L Berkowitz, \textit{Molecular dynamics simulation of a dipalmitoylphosphatidylcholine bilayer with NaCl}, Biophys. J. \textbf{84}, 3743 (2003).

\bibitem{pedersen:06} U R Pedersen, C Laidy, P Westh, G H Peters, \textit{The effect of calcium on the properties of charged phospholipid bilayers}, Biochim. Biophys. Acta \textbf{1758}, 573 (2006).

\bibitem{woolf:04}J N Sachs, H Nanda, H I Petrache, T B Woolf, \textit{Changes in phosphatidylcholine headgroup tilt and water order induced by monovalent salts: Molecular dynamics simulations}, Biophys. J. \textbf{86},  3772 (2004). 

\bibitem{mikami:07} K Shinoda, W Shinoda, M Mikami, \textit{Molecular dynamics simulation of an archeal lipid bilayer whit sodium chloride}, Phys. Chem. Chem. Phys. \textbf{9}, 643 (2007). 

\bibitem{yamada:05} N L Yamada, H Seto, T Takeda, M Nagao, Y Kawabata, K Inoue, \textit{SAXS, SANS and NSE studies on “unbound state” in DPPC/water/CaCl$_2$ system}, J. Phys. Soc. Jpn. \textbf{74}, 2853 (2005). 

\bibitem{smit:02} D Frenkel, B Smit, \textit{Understanding molecular simulations}, Academic Press, New York (2002).

\bibitem{jjlc:96a} J J L\'opez Cascales, J Garc\'ia de la Torre, S J Marrink, H J C Berendsen, \textit{Molecular dynamics simulation of a charged biological membrane}, J. Chem. Phys. \textbf{104}, 2713 (1996).

\bibitem{berendsen:90} W F van Gunsteren, H J C Berendsen, \textit{Computer simulations of molecular dynamics: Methodology, applications and perspectives in chemistry}, Angew. Chem Int. Ed. Engl. \textbf{29}, 992 (1990).
 

\bibitem{bockmann:03} R A  B\"ockmann, A  Hac, T Heimburg, H Grubm\"uller, \textit{Effect of sodium chloride on a lipid bilayer}, Biophys. J. \textbf{85}, 1647 (2003).

\bibitem{gurtovenko:08} A A Gurtovenko, I Vattulainen, \textit{Effect of NaCl and KCl on phosphatidylcholine and phosphatidylethanolamine lipid membranes: Insight from atomic-scale simulations for understanding salt-induced effects in the plasma membrane}, J. Phys. Chem. B. \textbf{112}, 1953 (2008). 

\bibitem{spce} H J C Berendsen, J R Grigera, T P Straatsma, \textit{The missing term in effective pair potentials}, J. Phys. Chem. \textbf{91}, 6269 (1987).

\bibitem{gromacs:95} H J C Berendsen, D van der Spoel, R van Drunen, \textit{A message-passing parallel molecular dynamics implementation}, Comp. Phys. Comm. \textbf{91}, 43 (1995).

\bibitem{gromacs:01} E Lindahl, B Hess, D van der Spoel, \textit{GROMACS 3.0: A package for molecular simulation and trajectory analysis}, J. Mol. Mod. \textbf{7}, 306 (2001).

\bibitem{darden:93} T Darden, D York, L Pedersen, \textit{Particle mesh Ewald: An N.log(N) method for Ewald sums in large systems},
J. Chem. Phys. \textbf{98}, 10089 (1993).

\bibitem{essmann:95} U Essmann, L Perea, M L Berkowitz, T Darden, H Lee, L G Pedersen, \textit{A smooth particle mesh Ewald method}, {J. Chem. Phys.} \textbf{103}, 8577 (1995).

\raggedbottom 
\pagebreak

\bibitem{young:88} L R De Young, K A Dill, \textit{Solute partitioning into lipid bilayer-membranes}, Biochemistry \textbf{27}, 5281 (1988).

\bibitem{seelig:74} A Seelig, J Seelig, \textit{The dynamic structure of fatty acyl chains in a phospholipid bilayer measured by deuterium magnetic resonance}, Biochemistry \textbf{13}, 4839 (1974). 

\bibitem{cevcbio81} G Cevc, A Watts, D Marsh, \textit{Titration of the phase transition of phosphatidilserine bilayer membranes. Effect of pH, surface electrostatics, ion binding and head-group hydration}, Biochemistry \textbf{20}, 4955 (1981). 

\bibitem{hebio82} H Hauser, F Paltauf, G G Shipley, \textit{Structure and thermotropic behavior of phosphatidylserine bilayer membranes}, Biochemistry \textbf{21}, 1061 (1982).


\bibitem{nagle:00} J F Nagle, S Tristam-Nagle, \textit{Structure of lipid bilayers}, Biochim. Biophy. Acta \textbf{1469}, 159 (2000).

\bibitem{lewis:83} B A Lewis, D M Engelman, \textit{Lipid bilayer thickness varies linearly with acyl chain length in fluid phosphatidylcholine vesicles}, J. Mol. Biol. \textbf{166}, 211 (1983).

\bibitem{pace:82} R J Pace, S I Cham, \textit{Molecular motions in lipid bilayer. I. Statistical mechanical model of acyl chains motion}, J. Chem. Phys. \textbf{76}, 4217 (1982).

\bibitem{thurmond:91} R L Thurmond, S W Dodd, M F Brown, \textit{Molecular areas of phospholipids as determined by 2H NMR spectroscopy}, Biphys. J. \textbf{59}, 108 (1991).

\bibitem{porasso:09} R D Porasso, J J L\'opez Cascales, \textit{Study of the effect of Na$^+$ and Ca$^{2+}$ ion concentration on the structure of an asymmetric DPPC/DPPS + DPPS lipid bilayer by molecular dynamics simulation}, Coll. and Surf. B. Bioint. \textbf{73}, 42 (2009).



\end{thebibliography}
\end{document}